# Chiral-Dependent Tensile Mechanics of Graphene

Young In Jhon[†,1] and Myung S. Jhon[*,1,2]

[1] Nano-convergence Core Technology for Human Interface (WCU), School of Advanced Materials Science and Engineering, Sungkyunkwan University, Suwon 440-746, Korea
[2] Department of Chemical Engineering, Carnegie Mellon University, Pittsburgh PA 15213, USA

**ABSTRACT:** We report a molecular dynamics study on the tensile mechanics of graphene as gradually rotating the tensile direction from armchair to zigzag direction, covering the complete range of chiral directions which has never been explored so far. We observed monotonic increases of tensile strength and strain as the chiral (rotational) angle increases. Key feature is their negligible changes up to chiral angle of ~12° and the subsequent rapid increases and this pattern holds for all temperatures examined here (100-700 K). Considering a topologically consistent (zigzag-lines) breaking of graphene, we presented a unified fracture model that successfully reproduced the simulation results as well as explaining their physical origin. Notably, we found that the elastic stress of graphene is quasi-isotropic for all chiral directions in contrast to its anisotropic fracture behavior. Through the indentation simulations of graphene, we demonstrated that our rationale established in uniaxial (1D) tensile systems is applicable to 2D tensile systems as well.

Graphene, a single layer of carbon atoms arranged in a honeycomb structure, has attracted a large amount of attentions due to its exceptional properties, such as superior electrical and thermal conductivities and an extraordinary mechanical strength.[1-7] It is thinnest and strongest than any other material ever discovered. Naturally, such superlative characteristics of graphene have triggered its active application to a wide range of engineering areas. For instance, there have been numerous attempts to improve mechanical strengths,[8-10] thermal properties,[11-13] and energy conversion efficiencies[14-16] of materials by incorporating graphene into them. Among the recent studies, one of the most featured applications would be the use of graphene for a support material of liquid cell in TEM study where the liquid was encapsulated by graphene membranes.[17] In that system, graphene played a critical role as enduring ultrahigh vacuum condition for a proper TEM operation as well as allowing very feasible electron transmission for high-resolution probing of the reaction occurring in the liquid, which eventually benefited from high tensile strength, flexibility, and ultrathin nature of this material. With growing demand on such graphene-based devices and hybrid materials, it becomes increasingly important to understand mechanical characteristics of graphene at a more in-depth level.

A number of studies have been performed so far to explore the mechanical fracture phenomena of ceramics and metals and their mechanisms can be well explained by Griffith's brittle fracture theory[18,19] and ductile fracture models for soft metals.[20,21] However, graphene is neither brittle nor ductile and essentially differs from either of them. This material is very strong and very stretchy at a time and its mechanical fracture behavior cannot be dictated by any single mechanism that ever existed.

Great efforts have been paid in characterizing the mechanical properties of graphene so far. By utilizing AFM-based technique, Lee *et al.* reported the Young's modulus as 1.0 7± 0.1 TPa and the intrinsic breaking strength as 130 ± 10 GPa, assuming the thickness of graphene to be 0.335 nm.[5]

On the other hand, most of previous studies for mechanical properties of graphene have been predominantly relying on computational methods such as *ab-initio* calculation,[22-24] tight-binding modeling,[25,26] and molecular dynamics simulations[27-31] due to the tremendous experimental difficulties for controlling a monoatomic film and developing appropriate measurement techniques. From these theoretical approaches, we were able to know many principal mechanical characteristics of graphene such as Poisson's ratio, temperature effect, and grain boundaries effect which had been extremely elusive via experimental approach.

Particularly, theoretical studies also suggested that graphene would exhibit the distinct anisotropic behavior in its tensile fracture, which is not experimentally observed yet. For instance, the molecular dynamics study indicated that the tensile strength and strain of graphene should be 107 GPa and 0.20 under the zigzag-directional elongation at 300 K while they should be 90 GPa and 0.13 under the armchair-directional elongation.[26] However, all precedent works on tensile mechanics of graphene were performed for the zigzag and armchair directional deformations only and it has never been inspected for the other chiral directions although graphene would be likely to deform along various directions in actual cases.

In this paper, to prevent such blind point, we have systematically studied the tensile mechanics of graphene using molecular dynamics simulations as gradually rotating the tensile direction from armchair to zigzag-direction. This rotational range covers the complete range of chiral directions of graphene due to its sixfold symmetry. Specifically, nine chiral directions were selected for this study as shown in Fig. 1 (a) and our tensile simulation system was briefly illustrated in Fig. 1 (b).

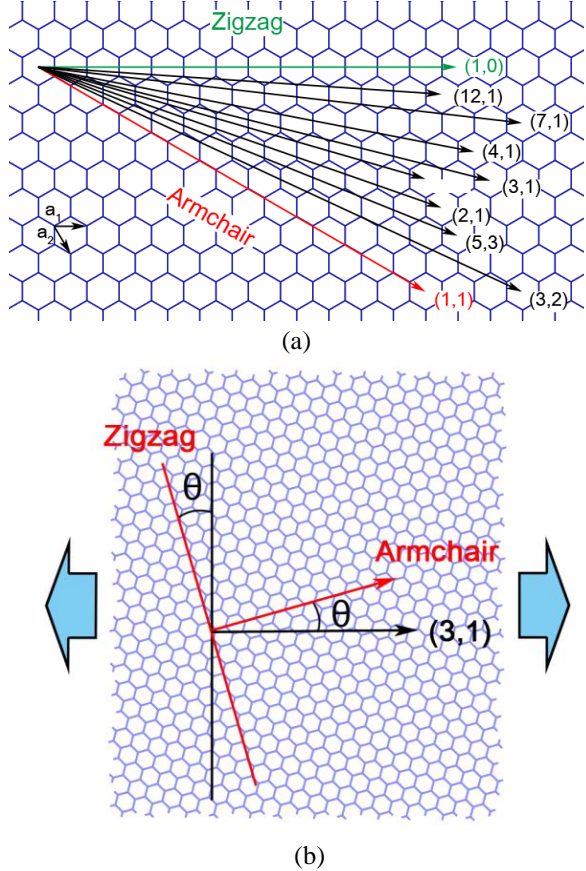

Fig. 1. (a) The nine chiral directions considered in this study to investigate chiral-dependent tensile mechanics of graphene. Their respective chiral notations are given in parenthesis. Any possible chiral direction would be located between armchair (red) and zigzag (green) directions due to the sixfold symmetry of graphene. (b) The schematic tensile simulation system in which the tensile direction was set to (3,1) chiral direction representatively. In this study, a chiral angle $\theta$ is defined as the angle made between the relevant tensile chiral direction and the armchair direction.

The molecular dynamics simulations were performed using the LAMMPS (Large-scale Atomic/Molecular Massively Parallel Simulator) software package[32] and the simulation systems were constructed in an orthogonal cell with the periodic boundary condition (Fig. 1 (b)). We employed the adaptive intermolecular reactive empirical bond order (AIREBO) potential[33] which has been widely used to study mechanical properties of carbon nanomaterials such as carbon nanotube, graphite, and graphene. The cut-off radius of the potential was set to be 2.0 Å to avoid spuriously high bond forces and unphysical results near the fracture region[27,29,34]. The dimension of the simulation system and atomic coordinates were first optimized using a gradient-based minimization method with tolerance criteria of $10^{-8}$ eV/Å in force and/or $10^{-8}$ eV in energy. Based on the system size obtained above, canonical NVT ensemble simulation was performed for $3\times10^5$ steps by increasing temperature gradually from zero to the desired temperature and consecutively equilibrated further for $7\times10^5$ steps under the isothermal condition. Then, the system was finally elongated with the strain rate of 0.0001 ps$^{-1}$ in the specific chiral direction where Non-equilibrium molecular dynamics (NEMD) simulation was employed to describe non-thermal streaming velocities of continuously strained system using SLLOD equations of motion coupled to Nose-Hoover thermostat[35].

The tensile strength of graphene was plotted as a function of chiral (rotation) angle $\theta$ in Fig. 2 where $\theta$ is defined to be the angle made between the armchair direction and the tensile direction (Fig. 1 (b)). Thus, for the armchair and zigzag tensile directions, $\theta$ should be 0° and 30°, respectively and $\theta$ would range between these values for an arbitrary tensile direction due to the sixfold symmetry of graphene. We found that key feature of the tensile strength chiral dependence is its negligible change up to chiral angle of ~12° and the subsequent rapid increase as raising the chiral angle. This pattern holds for all examined temperatures (namely, 100, 300, 500, and 700K) as shown in Figs. 2 (a)-(d), although the tensile strength quantitatively decreased as temperature increased due to thermal softening. For the tensile fracture strain, we observed the very similar chiral dependence to that of the tensile strength (Fig. 3).

To see their chiral-dependence more manifestly excluding the thermal effect, we also plotted the tensile strength and strain in a reduced form by dividing them with their values for chiral angle of 0° at the respective temperatures. Their curves showed an excellent coincidence each other when they were put together (Fig. S1), which indicates that an identical physical origin may exist behind these phenomena regardless of the temperature.

To gain insight into the physics behind them, we carefully monitored the evolution of their atomic structures during the tensile process. We found that the fracture of graphene dominantly occurred along zigzag-lines of the hexagonal structure regardless of the deformational direction and temperature (Fig. S2). The similar results were reported previously in the axial elongation of carbon nanotubes and the longitudinal elongation of graphene nanoribbon.[27,36] We speculated that a feasible breaking of graphene occurring along zigzag-lines much resembled the slip or breaking of metals which takes place on the densest plane of the crystal structure under the tensile deformation. Based on this idea, the stress transformation formalism[37] was adopted for evaluating the tensile strength of graphene for an arbitrary tensile direction as given by

$$S_n = \frac{1}{2}(S_x + S_y) + \frac{1}{2}(S_x - S_y)\cos 2\theta + T_{xy}\sin 2\theta \quad (1)$$

where the $S_n$ is the normal (engineering) stress of an arbitrary plane (P) that is perpendicular to the x-y plane, $\theta$ is an angle made between x-direction and the normal direction of P-plane, and $S_x$, $S_y$, and $T_{xy}$ refer to the corresponding components of Cauchy stresses, respectively.

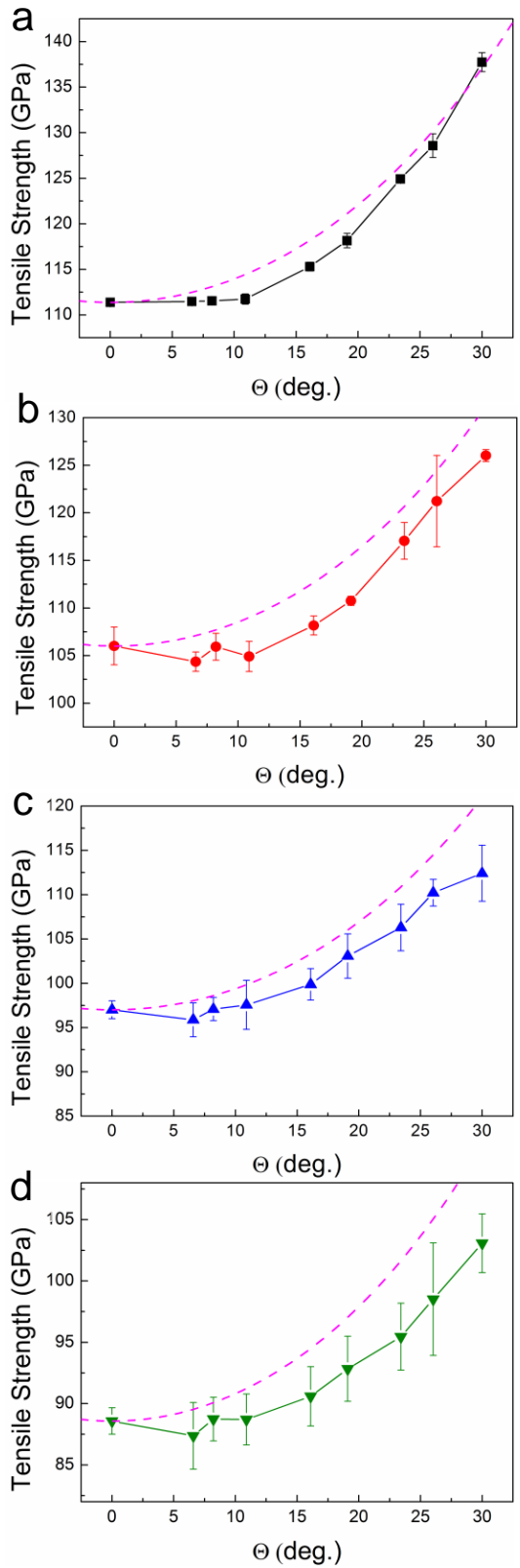

Fig. 2. Tensile strength of graphene plotted as a function of the chiral angle at (a) 100 K, (b) 300 K, (c) 500 K, and (d) 700 K, respectively. The dashed line indicates the theoretical values obtained from our unified fracture model.

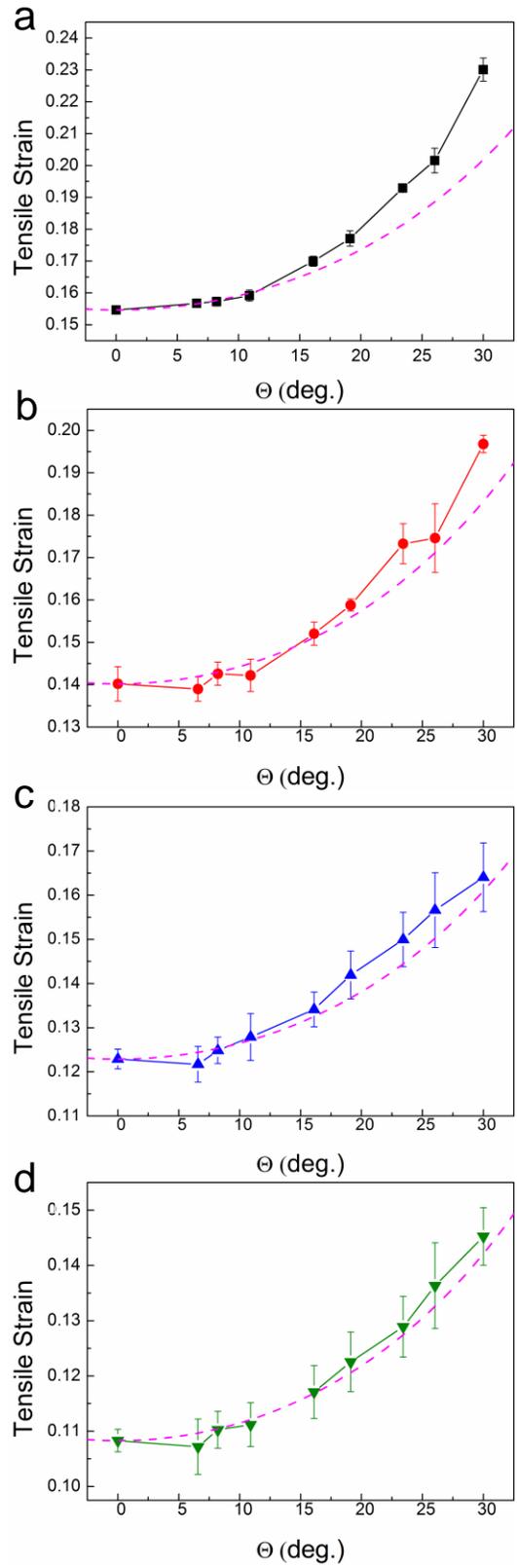

Fig. 3. Tensile fracture strain of graphene plotted as a function of the chiral angle at (a) 100 K, (b) 300 K, (c) 500 K, and (d) 700 K, respectively. The dashed line indicates the theoretical values obtained from our unified fracture model.

We assumed three things in this equation for the application to the uniaxial tensile fracture mechanics of graphene. First, $x$-direction and normal direction $n$ denote the tensile direction and the armchair-direction that is closest to the tensile direction, respectively. Second, $S_y$ and $T_{xy}$ are negligible during the tensile process. Third, the elongation is completed to the fracture point. From these assumptions, we obtained the equation given by

$$S_{AC} = \frac{S_{tensile}}{\cos^2 \theta'} \quad (2)$$

where $S_{AC}$ is tensile strength in armchair direction; $S_{tensile}$ is tensile strength in an arbitrary tensile direction; and $\theta'$ is the angle made between the armchair direction and the tensile direction (Fig. 4). If we assume the brittle fracture of graphene, $\theta'$ would be approximated to be the chiral angle of original grahene structure (namely, $\theta$ in Fig. 4 (b)). In this case, we may also use Cauchy-Born rule[36] to evaluate the tensile strain of graphene. However, graphene is not brittle and it is considerably stretchable before the fracture occurs. In addition, the effective magnitude of $S_{AC}$ will be also remarkably reduced as the chiral angle increases, due to the transverse stretching effect. Taking these factors into account, we assumed that $\theta'$ of Eq. (2) should be the angle measured at the fracture point rather than that of the original structure (Fig. 4(b)).

For this regime, by inspecting the geometric relationship made during the tensile process (Fig. 4), we derived the equations for several significant (engineering) strains as written by

$$e_{tensile} + 1 = \frac{L'}{L} = \frac{\tan \theta}{\tan \theta'} = \frac{\tan \theta''}{\tan \theta} \quad (3)$$

$$e_{AC} + 1 = \frac{L_{AC}'}{L_{AC}} = \frac{\sin \theta}{\sin \theta'} \quad (4)$$

$$e_{ZZ} + 1 = \frac{L_{ZZ}'}{L_{ZZ}} = \frac{\cos \theta}{\cos \theta''} \quad (5)$$

From Eqs (3)-(5), we can finally evaluate the values for $e_{tensile}$ and $\theta'$ (eventually, the value of $S_{tensile}$ using Eq. (2)) since $e_{AC}$, $S_{AC}$, and $\theta$ are known. Here, the contraction transverse to the tensile direction was not considered as such in most of the bulk system tensile simulations.

The theoretical values obtained from our aforementioned fracture model were given in dashed magenta lines in Figs. 3 and 4, showing a good coincidence with the simulation results while a model based on the brittle fracture yielded remarkably difference between them (Fig. S3).

Besides the study on chiral-dependent tensile fracture mechanics, we have also investigated how the tensile elastic behavior of graphene varied according the change of the tensile direction. Notably, we observed that the tensile elastic stress of graphene was almost the same for the different tensile directions as far as it was elongated to the same strain (Fig. 5).

It indicates that graphene would exhibit a quasi-isotropic behavior in the elastic region in contrast to its anisotropic response to the fracture strength. It was previously reported that graphene showed similar elastic behaviors for armchair and zigzag directions[27], however, we were not able to assure that its elastic behavior is indeed quasi-isotropic since it was never examined for other chiral directions. In this context, our study is the first to prove the (quasi) isotropic behavior of graphene for the elastic tensile motion. We found that this pattern held for all examined temperatures (Fig. S4).

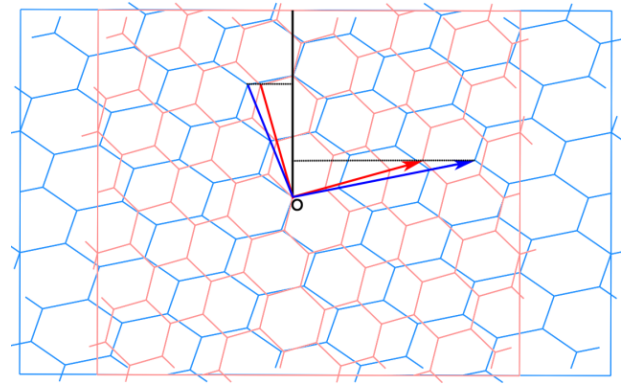

(a)

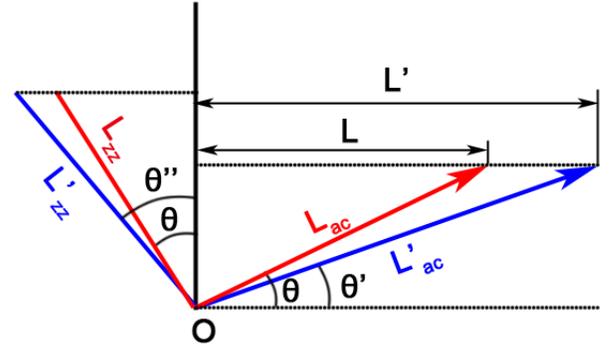

(b)

Fig. 4. (a) The geometric variation of graphene occurring in the uniaxial tensile process. (b) The geometric diagram relating the angles of $\theta$, $\theta'$, and $\theta''$ with the lengths along armchair, zigzag, and tensile directions of graphene. The red line and arrow indicate the lengths in armchair and zigzag directions in the original graphene structure, respectively, while the blue line and arrow indicate those in its elongated structure.

In Lee et al.'s AFM study[5], isotropic mechanical response of graphene was assumed based on the sixfold rotation symmetry in deriving the equation which relates the tensile strength magnitudes and the measured values of indenter force. Regarding this point, we suggested that such isotropic mechanical response of graphene would be realized at a more fundamental level, validating Lee et al.'s assumption more definitely.

For the further validation of our rationale on tensile mechanics of graphene, we also performed the indentation simulations of graphene (Fig. 6 (a)). This indentation process belongs to two-dimensional tensile system differently from uniaxial elongation process and thus, it would be interesting to see how graphene responses to this process at the aspects of its anisotropic/isotropic mechanical behaviors. Details of our method were similar to those of Lee *et al.*'s experimental study except for the shape of wells patterned in SiO$_2$ substrate. We drilled ellipsoidal wells instead of spherical ones in which the long-axis of the well was oriented to either of armchair and zigzag directions to see the possible anisotropic effect (Fig. 6 (b)).

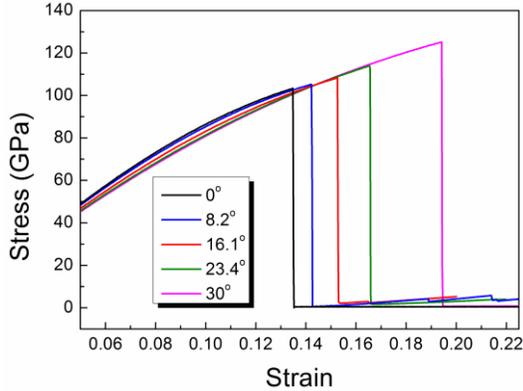

Fig. 5. The tensile stress-strain curves of graphene obtained for different tensile directions the chiral angles of which were given in the box. They showed very similar nonlinear responses to each other in the elastic region regardless of their respective tensile directions suggesting a quasi-isotropic tensile elastic behavior of graphene. In this figure, we only presented the result obtained at 300 K but this pattern held for 100, 500, and 700 K as well.

Ellipsoidal wells (the diameters of long and short axes were 120 Å and 60 Å, respectively) were patterned onto a SiO$_2$ substrate (depth ~10 Å) for every closely-packed rectangular SiO$_2$ domain of 138 Å by 136 Å. Then, a graphene film was subsequently attached onto the substrate and the indenter was initially positioned 8 Å above the graphene film at the center of the well (Fig. 6 (a)). Finally, the indenter moved down at the rate of 0.5 Å/ps for indenting the graphene film in which the atoms of two utmost bottom layers of SiO$_2$ substrate were fixed and the thermal motion of the indenter was not considered. The system temperature was set to be 100 K and details on atomic potentials employed in our indentation simulations are given in supplementary information. The diameter of the indenter was around 20 Å and the force exerted on each atom due to the indenter was given by

$$F(r) = K \times (r - R)^2 \quad (6)$$

where $K$ is the specified force constant, $r$ is the distance from each atom to the center of the indenter, and $R$ is the radius of the indenter. The force is repulsive and $F(r) = 0$ for $r > R$. Here, $K$ and $R$ were set to be 1000 eV/Å$^3$ and 10 Å, respectively.

As gradually moving the indenter downward, we measured the force exerted on the indenter for the two systems having differently oriented ellipsoidal wells (Fig. 6 (c)). Basically, we obtained the same result to that of uniaxial tensile process. They showed very similar elastic behaviors quantitatively and qualitatively, regardless of the well orientation. It means that our conclusion for elastic tensile behavior of graphene, established in its uniaxial (1D) tensile mechanics, is still operative for the two-dimensional (2D) tensile deformation.

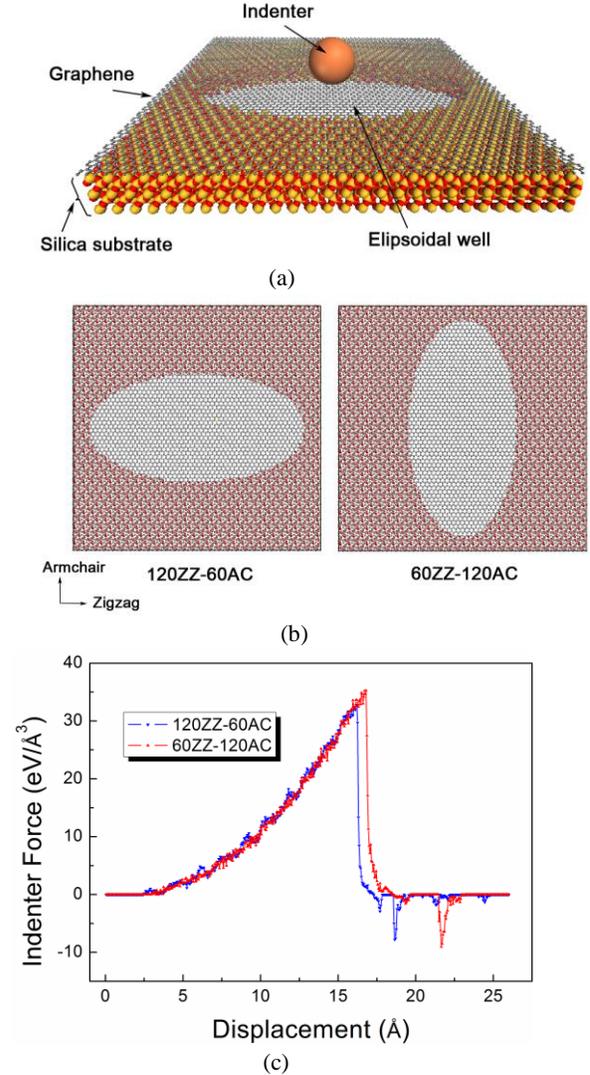

Fig. 6. (a) The indentation simulation system employed in the study for two-dimensional tensile deformation. (b) Two differently oriented ellipsoidal wells that are patterned on the silica substrate. (c) The measured values of the force exerted on the indenter as it gradually moves down for the indentation of the graphene film.

In contrast to their isotropic elastic responses, these systems yielded distinctly different values for the maximum indenter force (*i.e.*, the breaking force) and its corresponding indenter position. When the longer axis of the ellipsoidal well was aligned to the armchair direction (denoted as 60ZZ-120AC), it required larger amounts of breaking force and movement of the indenter (34.71 ± 0.48 eV/Å and 16.60 ± 0.13 Å, respectively) compared to those (32.72 ± 0.59 eV/Å and 16.09 ± 0.11 Å, respectively) in its configuration counterpart (denoted as 120ZZ-60AC). This anisotropic difference can be explained as follows. The effective tensile strain along the shorter axis would be greater than that of longer axis for the identical movement of the indenter. In addition, the fracture of graphene is critically determined by the deformation along the weakest tensile direction, *i.e.*, the armchair direction. In this context, 60ZZ-120AC is expected to be stronger than 120ZZ-60AC, which coincides with the simulation result indeed.

Such anisotropic behavior agrees well with our understanding for uniaxial tensile fracture of graphene. However, the anisotropic effect observed in the indentation process was much smaller than that of the uniaxial tensile process. On the basis of a continuum model for linear elastic, circular membrane under a spherical indenter[38], the tensile strengths of 60ZZ-120AC and 120ZZ-60AC were estimated to be 116.15 GPa and 112.78 GPa (these values are within reasonable range considering the uniaxial tensile simulation result), respectively. We attributed this attenuation effect to the fact that the effective tensile strength was averaged over all chiral directions in the indentation (2D) case. More another important factor would be that many of its constituent stresses (envision the disassembly of the stress to each chiral-directional component) possessed an elastic characteristic (isotropic) rather than the tensile strength feature (anisotropic) when the breaking started.

In summary, as gradually varying the tensile direction, we have systematically investigated chiral-dependent tensile mechanics of graphenene using molecular dynamics simulations. Any (tensile) chiral direction in graphene should be placed between the armchair and zigzag directions due to the sixfold rotation symmetry. Thus, after defining the chiral angle as the angle made between a certain chiral direction and the armchair direction, tensile simulations were performed by gradually varying the chiral angle from 0° to 30°, covering the complete range of possible tensile directions. For the tensile strength, key feature was its negligible change up to chiral angle of ~12° and the subsequent rapid increase as the chiral angle increased. This pattern held for all temperatures examined here (100-700 K) except for some quantitative lowering of the strength due to thermal softening. A very similar chiral dependence was observed in evaluating the fracture strain. Assuming a topologically consistent breaking in graphene, we presented a unified fracture model which successfully reproduced the simulation results. Notably, we found that the tensile elastic behavior of graphene was almost unaffected by the change of tensile direction in contrast to its anisotropic fracture response, suggesting the quasi-isotropic elastic behavior of graphene. Finally, our rationale for uniaxial tensile (1D) mechanics of graphene was revalidated and extended to two-dimensional cases through the indentation simulations. We believed that the features explored here will contribute significantly to understanding tensile mechanics of graphene at the in-depth level and have important implication in graphene-based devices and hybrid materials.

## Author information

**Corresponding Authors**

† E-mail: yijhon@kaist.ac.kr
* E-mail: mj3a@andrew.cmu.edu

## Acknowledgement

This work was supported by the World Class University program of KOSEF (Grant No. R32-2008-000-10124-0).

Supporting Online Material for

# Chiral-Dependent Tensile Mechanics of Graphene

Young In Jhon[1] and Myung S. Jhon[1,2]

[1] Nano-convergence Core Technology for Human Interface (WCU), School of Advanced Materials Science and Engineering, Sungkyunkwan University, Suwon 440-746, Korea

[2] Department of Chemical Engineering, Carnegie Mellon University, Pittsburgh PA 15213, USA

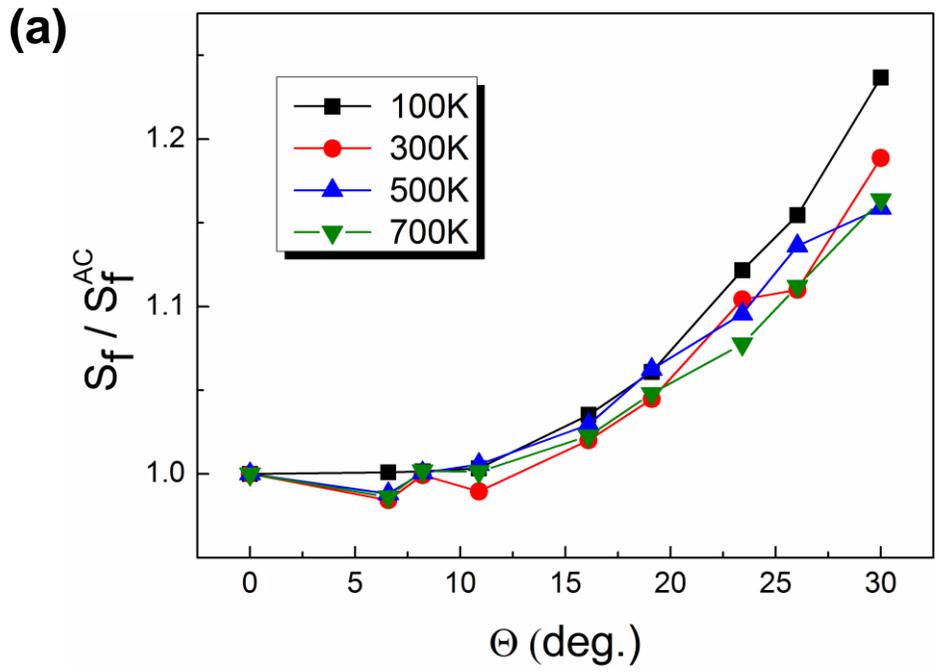

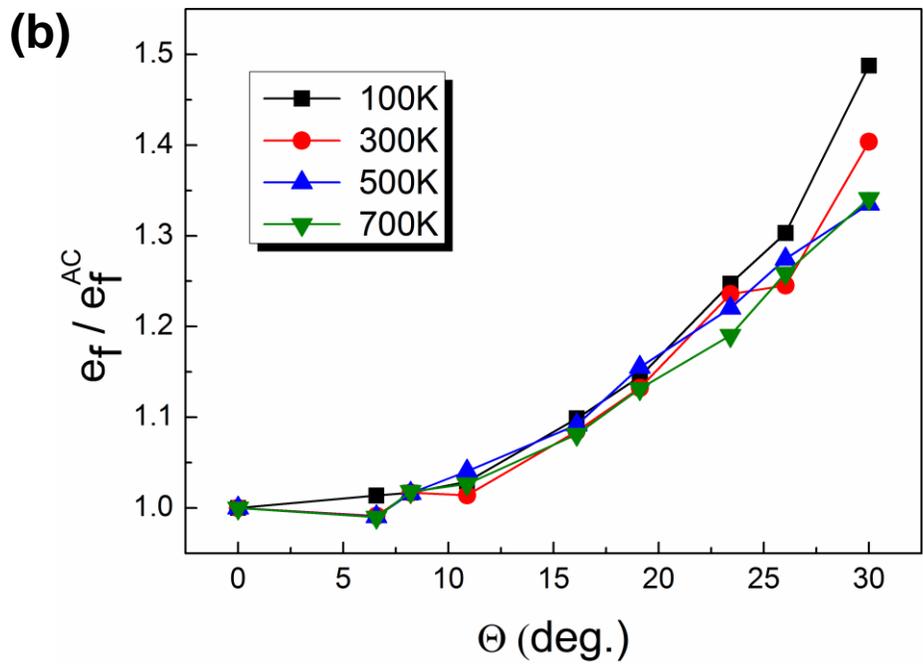

**Figure S1.** (a) The tensile strength and (b) the fracture strain of graphene plotted as a function of a chiral angle. They were shown in a reduced form by dividing their respective values for chiral angle 0° at various temperatures.

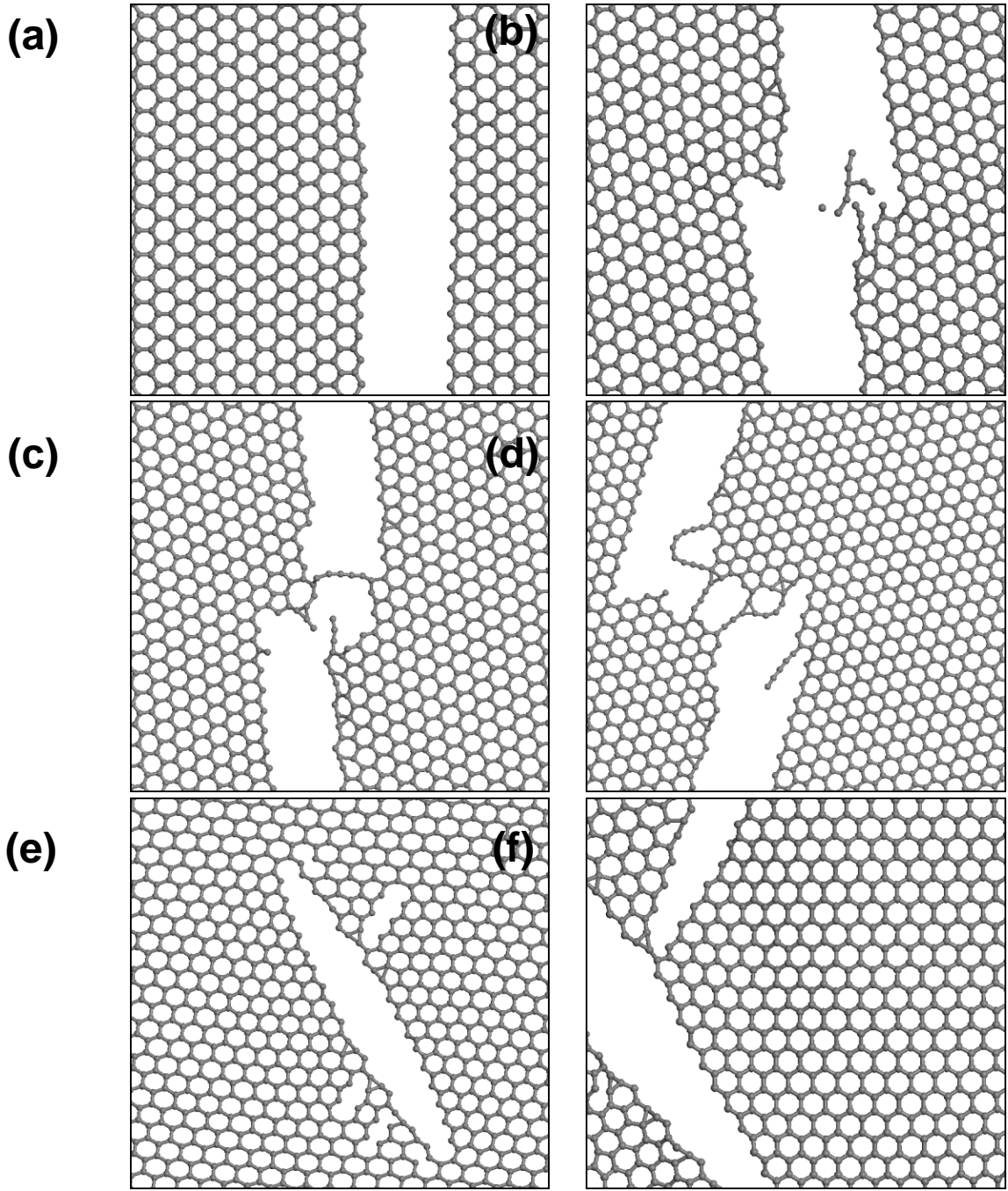

**Figure S2.** The structures appearing at the tensile fracture point of graphene for the different tensile (chiral) directions of (a) (1,1), (b) (3,2), (c) (5,3), (d) (3,1), (e) (7,1) and (f) (1,0). Their corresponding chiral angles are 0°, 6.6°, 8.2°, 16.1°, 23.4°, and 30°, re-

spectively and we see that graphene always breaks along zigzag-lines of its hexagonal structure for all tensile directions.

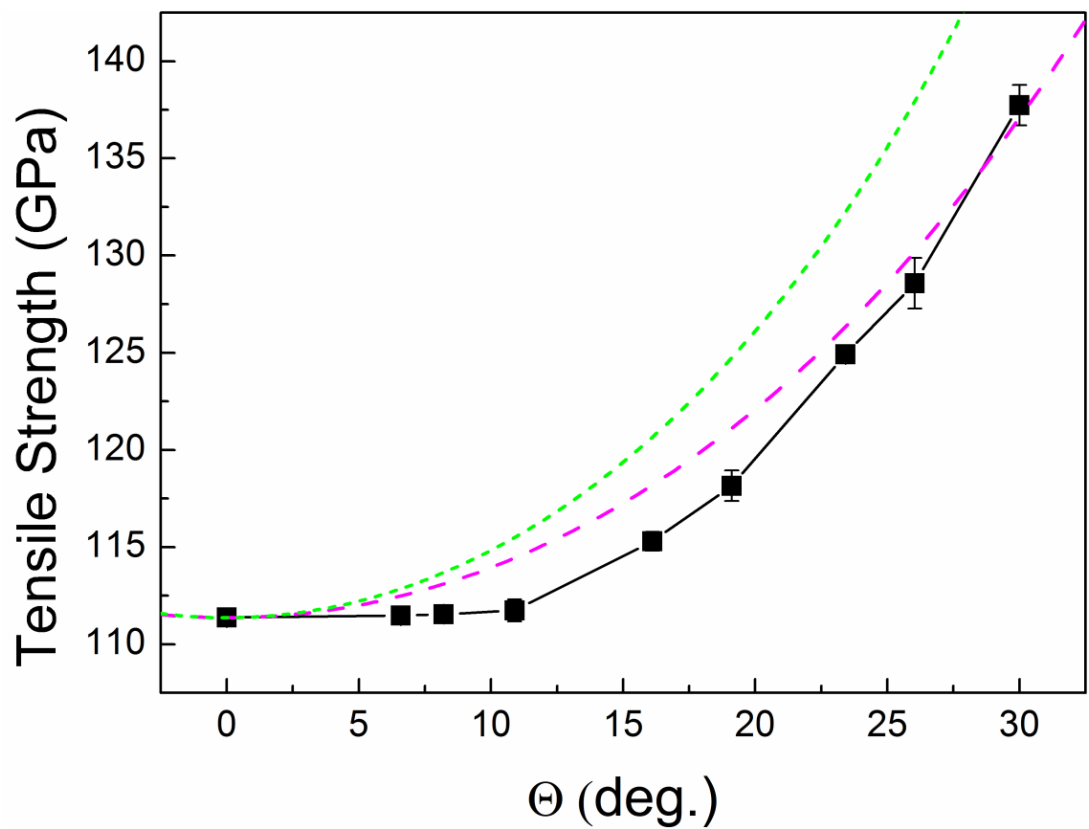

**Figure S3.** The tensile strength of graphene as a function of chiral angle at 100 K. The magenta dashed line indicates the results obtained from our unified fracture model while the green short-dashed line indicates the results obtained from the same fracture model but employing the (incorrect) brittle fracture assumption. We see that the former

is in a good agreement with the molecular dynamics simulation result (black square and line) while the latter is much deviated from the simulation result compared to the former.

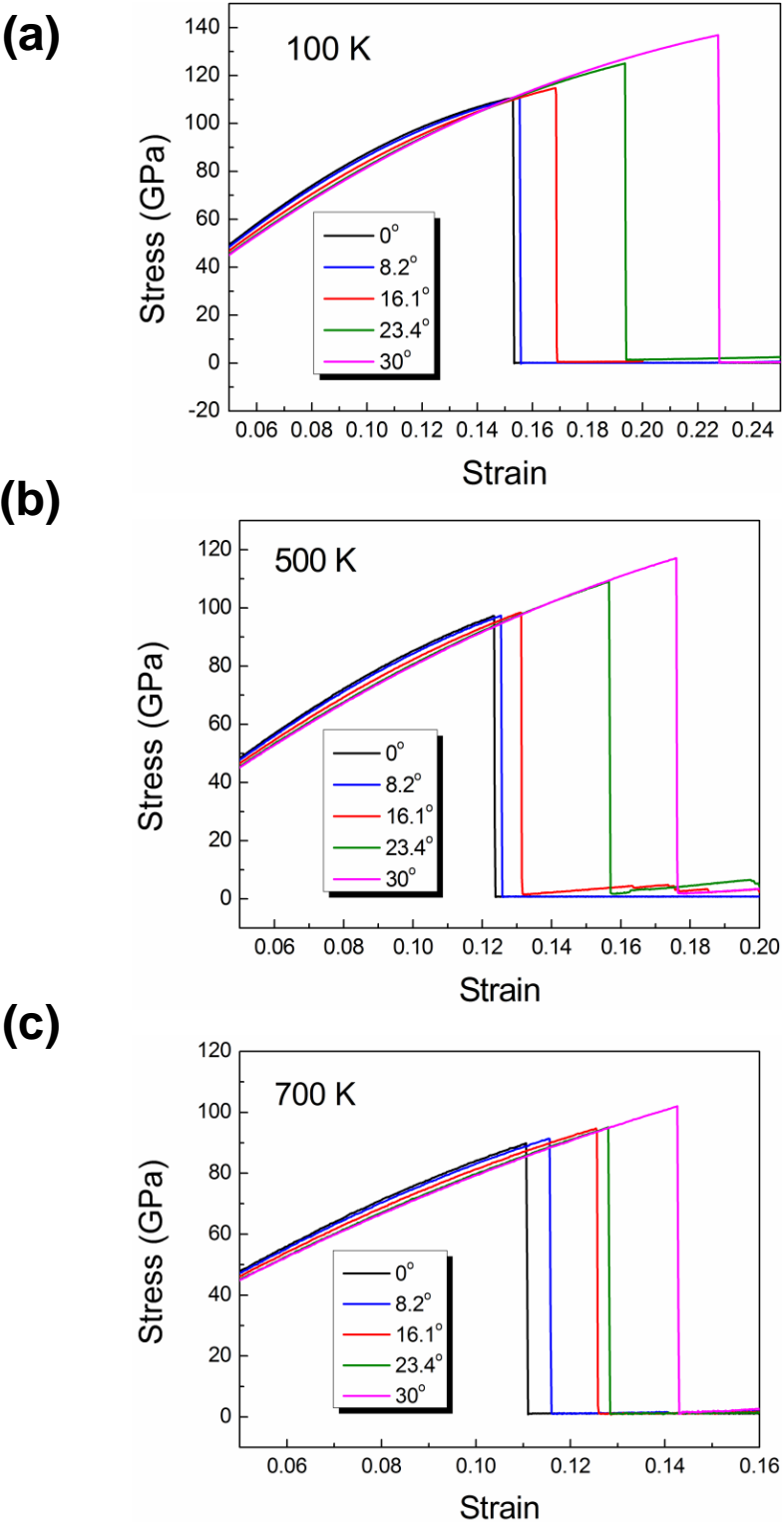

**Figure S4.** The stress vs. strain of graphene under the uniaxial elongation for different tensile directions at the temperatures of (a) 100K, (b) 500 K, and (c) 700 K.

## Atomic Potentials for Indentation MD Simulations

1. The interaction between carbon atoms of graphene:
**AIREBO (Adaptive Intermolecular Reactive Bond Order) potential**[1]
2. The interaction potential between silicon and oxygen atoms of $SiO_2$ substrate: **Tersoff potential**[2]
3. The (non-bonded) interaction between carbon atoms of graphene and silicon & oxygen atoms of $SiO_2$ substrate:
**Lenard-Jones potential**

$$E = 4\varepsilon\left[\left(\frac{\sigma}{r}\right)^{12} - \left(\frac{\sigma}{r}\right)^{6}\right], \ r < r_c$$

where the parameter values for each atomic pair are obtained using a mixing rule and they are shown in table S1 below.

Table S1. The parameter values for Lenard-Jones potentials between carbon & silicon and carbon & oxygen atoms.

| Atomic Pair | ε (eV) | σ (Å) | $r_c$ (Å) |
|---|---|---|---|
| Carbon & Silicon | 0.008909 | 3.326 | 8.315 |
| Carbon & Oxygen | 0.003442 | 3.001 | 7.503 |